# The thermal effect on the left-handedness of the mesoscopic composite right-Left handed transmission line

Xiao-Jing Wei, Shun-Cai Zhao* [1], Hong-Wei Guo

[a]Department of Physics, Faculty of Science, Kunming University of Science and Technology, Kunming, 650500, PR China

**Abstract**

Starting from the quantum fluctuation of current in the mesoscopic composite right-left handed transmission line (CRLH-TL) in the thermal Fock state, we investigate the left-handedness dependent of the frequencies, intensity and quantum fluctuations of the current field in the CRLH-TL under different thermal environment. The results show that the intensity and quantum fluctuations of current field in lower frequency bands affect the left-handedness distinctly under different thermal environment. The thermal effect on the left-handedness in the mesoscopic CRLH-TL deserves further experimental investigation in its miniaturization application.

## 1. Introduction

Left-handedness indicates materials exhibiting simultaneously negative effective permittivity $\epsilon$ and permeability $\mu$ in the same frequency window [1, 2, 3], which has been the center of significant interest in the physics and engineering communities since it was first realized at microwave frequencies [4] due to their exotic electromagnetic properties and their fascinating applications. However, one key obstacle in the applications of left-handed materials is absorption which is particularly important in the optical regime[5, 6, 7], and significant efforts have been spent to realize low-loss left-handed materials[8, 9, 10]. Meanwhile, the transmission lines (TLs) are reportedly low dissipation and wide frequency bands for left-handed materials[11, 12, 13, 14], i.e., the composite right-Left handed transmission lines (CRLH-TLs)[15]. Recently, an universal applications for CRLH-TLs are implemented by its left-handed phenomena. Such as in the acoustic wave frequency band, one-dimensional acoustic negative refractive index metamaterial is presented by the CRLH-TL[16], and the CRLH-TL is firstly reported to achieve the acoustic dispersive prism which has the capability of splitting a broadband acoustic wave into its constituent Fourier components within the audible frequency range of 800 Hz-1300 Hz[17]. In the the antenna application, the fractional bandwidth can be enhanced from 0.31% to 12.5% in a resonant antenna with an asymmetric coplanar waveguide based on the CRLH-TL[18]. And a new class of miniaturized nonreciprocal leaky-wave antenna is proposed for miniaturization, nonreciprocal properties and wide-angle scanning at the same time[19]. With four unit cells of CRLH-TL a wide-band loop antenna is proposed in a compact size[20]. Nowadays, under the nanotechnology and microelectronics influence, the compact application has turned to a tendency for CRLH-TLs.

---

[1]Corresponding author: zhaosc@kmust.edu.cn.

However, when the compact size of the CRLH-TL approaches the Fermi wavelength, the quantum effects on the CRLH-TL must be taken into account similarly to the mesoscopic circuits[21, 22, 23, 24, 25]. In our former work, we firstly deduced the quantum features of negative refraction index (NRI) of the lossless mesoscopic left-handed transmission line (LH-TL)[26]. And then the quantized lossy LH-TL[27] entered our the next work, we discussed the characteristics of NRI of the lossy LH-TL[27] in a displaced squeezed Fock state. And some novel quantum effects caused by the dissipation were revealed and were reputed significance for the miniaturization application of LH-TL.

Thus, from the point of against this miniaturization application challenge, this paper exploits the thermal effect on the left-handedness of the CRLH-TL in the thermal Fock state. And this paper is organized as follows. In Sec.2, we quantize the travelling current field in the unit-cell circuit of the CRLH-TL, and deduce the permittivity $\epsilon$ and permeability $\mu$ in the thermal Fock state. Then we evaluate the left-handedness in the thermal Fock state in Sec.3. Sec.4 presents our summary and conclusions.

## 2. The quantized CRLH-TL in the thermal Fock state

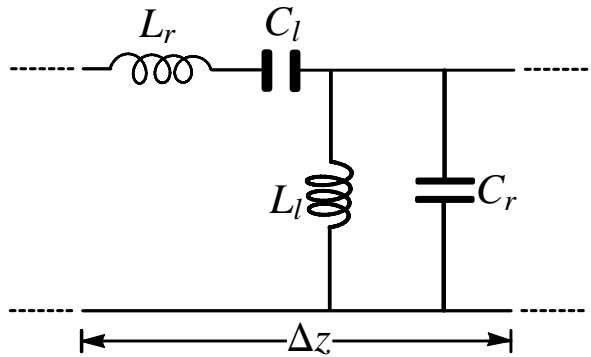

Figure 1: Schematic diagram of equivalent unit-cell circuit of the mesoscopic CRLH-TL.

The equivalent unit-cell circuit model of the proposed CRLH-TL is shown in Fig. 1. It consists of the series capacitor $C_l$ and inductance $L_r$, shunt inductance $L_l$ and capacitor $C_r$[28]. And the dimension $\Delta z$ of the equivalent unit-cell circuit is much less than the wavelength at operating frequency. The permittivity $\epsilon$, permeability $\mu$ for the unit-cell circuit in Fig.1 read as $\epsilon = C_r - \frac{1}{\omega^2 L_l}$, $\mu = L_r - \frac{1}{\omega^2 C_l}$ [29], respectively. Hence, let us now consider Kirchhoffs voltage and current laws for the unit-cell circuit in Fig. 1, which respectively read

$$u(z,t) = i(z,t)[\frac{1}{j\omega \frac{C_l}{\Delta z}} + j\omega L_r \Delta z] + u(z+\Delta z, t),$$

$$i(z,t) = [\frac{1}{j\omega \frac{L_l}{\Delta z}} + j\omega \frac{C_r}{\Delta z}]u(z+\Delta z, t) + i(z+\Delta z, t)$$

where $u(z,t)$ is the voltage, $i(z,t)$ is the current, and $\omega$ is the angle frequency. When $\Delta z \to 0$, the above equations, lead to the following system:

$$\frac{\partial^2 u(z,t)}{\partial z^2} = -[\omega L_r - \frac{1}{\omega C_l}] \times [\omega C_r - \frac{1}{\omega L_l}]u(z,t),$$

$$\frac{\partial^2 i(z,t)}{\partial z^2} = -[\omega L_r - \frac{1}{\omega C_l}] \times [\omega C_r - \frac{1}{\omega L_l}]i(z,t)$$

The above equations, together with the auxiliary equations $i(z,t) = C\frac{\partial u(z,t)}{\partial t}$ and $C = C_r + C_l$ lead to the forward plane-wave solutions:

$$u(z,t) = Ae^{-j(wt-\beta z)} + A^* e^{j(wt-\beta z)} \tag{1}$$

$$i(z,t) = jwC[A^* e^{j(wt-\beta z)} - Ae^{-j(wt-\beta z)}] \tag{2}$$

with $\beta = \sqrt{(\frac{\omega}{\omega_r})^2 + (\frac{\omega_l}{\omega})^2 - k\omega_l^2}$, where $\omega_r = \frac{1}{\sqrt{L_r C_r}}$, $\omega_l = \frac{1}{\sqrt{L_l C_l}}$, $k = L_r C_l + L_l C_r$. In Eq.(1) and Eq.(2), $A^*$ is the complex conjugate of $A$ for the normalization purpose. To simplify Eq.(1) and Eq.(2), two functions $\xi(z,t)$ and $\eta(z,t)$ are introduced by

$$u(z,t) = \xi(z,t),$$

$$i(z,t) = \frac{\omega^2}{z_0}\eta(z,t)$$

where $z_0$ is the length of per unit circuit. We assume that $\omega$ is given and choose the unit length of circuit $z_0$ to be fixed. Then, $\eta(z,t)$ and $\zeta(z,t)$ differential $t$ are

$$\frac{\partial \xi(z,t)}{\partial t} = \frac{\omega^2}{Cz_0}\eta(z,t), \tag{3}$$

$$\frac{\partial \eta(z,t)}{\partial t} = -Cz_0 \xi(z,t) \tag{4}$$

It is straightforward to find the following relation,

$$\frac{\partial}{\partial \xi(z,t)}\left(\frac{\partial \xi(z,t)}{\partial t}\right) + \frac{\partial}{\partial \eta(z,t)}\left(\frac{\partial \eta(z,t)}{\partial t}\right) = 0 \qquad (5)$$

So, $\eta(z,t)$ and $\zeta(z,t)$ are the canonically conjugate variables and obey the hamiltonian canonical equations,

$$\frac{\partial \xi(z,t)}{\partial t} = \frac{\partial H}{\partial \eta(z,t)}, \qquad (6)$$

$$\frac{\partial \eta(z,t)}{\partial t} = -\frac{\partial H}{\partial \xi(z,t)} \qquad (7)$$

The integrals of Eq.(6) and Eq.(7) can deduce the hamiltonian for the electromagnetic wave with the ignored integral initial constant as follows,

$$H = \frac{\omega^2}{2Cz_0}\eta^2(z,t) + \frac{1}{2}Cz_0\xi^2(z,t) \qquad (8)$$

If we associate hermitian operators of $\hat{\eta}(z,t)$ and $\hat{\xi}(z,t)$ and require that they satisfy the commutation relation,

$$[\hat{\xi}, \hat{\eta}] = j\frac{Cz_0}{\omega}[Ae^{-j(wt-\beta z)} + A^*e^{j(wt-\beta z)}, A^*e^{j(wt-\beta z)} - Ae^{-j(wt-\beta z)}]$$
$$= j\frac{2Cz_0}{\omega}[\hat{A}, \hat{A}^*]$$
$$= j\hbar$$

then, we achieve the quantized unit-cell circuit with the definitions,

$$\hat{A} = \hat{a}\sqrt{\frac{\hbar\omega}{2Cz_0}}$$

$$\hat{A}^* = \hat{a}^*\sqrt{\frac{\hbar\omega}{2Cz_0}}$$

Then the two operators $\hat{\eta}(z,t)$ and $\hat{\xi}(z,t)$ are,

$$\hat{\xi}(z,t) = \sqrt{\frac{\hbar\omega}{2Cz_0}}[\hat{a}e^{-j(wt-\beta z)} + \hat{a}^\dagger e^{j(wt-\beta z)}], \qquad (9)$$

$$\hat{\eta}(z,t) = j\sqrt{\frac{\hbar\omega}{2Cz_0}}[\hat{a}^\dagger e^{j(wt-\beta z)} - \hat{a}e^{-j(wt-\beta z)}] \qquad (10)$$

and the quantum Hamiltonian of the unit-cell circui can be written as $\hat{H} = \hbar\omega(\hat{a}^\dagger\hat{a} + \frac{1}{2})$ which is analogous to the oscillator's Hamiltonian operator in Schrödinger representation.

As for the equilibrium situation, the so-called thermo field dynamics (TFD) extends the usual quantum field theory to a finite temperature[30], in which the tilde space accompanies with the Hilbert space. The creation and annihilation operators $\hat{a}^\dagger$, $\hat{a}$ associate with their tilde operators $\tilde{\hat{a}}^\dagger$, $\tilde{\hat{a}}$ according the rules[31]: $[\tilde{\hat{a}}, \tilde{\hat{a}}^\dagger] = 1, [\tilde{\hat{a}}, \hat{a}] = [\tilde{\hat{a}}, \hat{a}^\dagger] = [\hat{a}, \tilde{\hat{a}}^\dagger] = 0$. The number operators in the Hilbert space and tilde space are read as $\hat{n} = \hat{a}^\dagger \hat{a}$, $\tilde{\hat{n}} = \tilde{\hat{a}}^\dagger \tilde{\hat{a}}$. In this direct product space, the thermal Fock state at finite temperature $|\hat{n}\tilde{\hat{n}}\rangle_T$ can be built by the thermal Bogoliubov transformation[31] through the Fock state at zero temperature $|\hat{n}\tilde{\hat{n}}\rangle$ : $|\hat{n}\tilde{\hat{n}}\rangle_T = \hat{T}(\theta)|\hat{n}\tilde{\hat{n}}\rangle$, where $\hat{T}(\theta)$ is a thermal unitary operator which is defined as

$$\hat{T}(\theta) = exp[-\theta(\hat{a}\tilde{\hat{a}} - \hat{a}^\dagger\tilde{\hat{a}}^\dagger)] \tag{11}$$

the parameter $\theta$ is the thermal parameter relating the thermal photos $n_0$ in the thermal vacuum state: $n_0 = sinh\theta$. The thermal photos $n_0$ and temperature $T$ are ruled by the Boltzmann distribution $n_0 = [exp(\hbar\omega/k_B T) - 1]^{-1}$, in which $k_B$ is the Boltzmann constant.

Then the bosonic operators in TFD can relate each other by the thermal Bogoliubov transformation as following,

$$\hat{T}^\dagger(\theta)\hat{a}\hat{T}(\theta) = \mu\hat{a} + \tau\tilde{\hat{a}}^\dagger, \tag{12}$$

$$\hat{T}^\dagger(\theta)\hat{a}^\dagger\hat{T}(\theta) = \mu\hat{a}^\dagger + \tau\tilde{\hat{a}} \tag{13}$$

where $\mu = cosh\theta$, $\tau = sinh\theta$. With the quantum fluctuation of the current $\langle(\Delta i)^2\rangle = (\langle\hat{i}^2\rangle - \langle\hat{i}\rangle^2)$ in the thermal Fock state in Heisenberg picture and the definitions of effective permittivity $\epsilon$, permeability $\mu$ for the unit-cell circuit [29], we deduce the permittivity $\epsilon$, permeability $\mu$ in the thermal Fock state as follows,

$$\epsilon = C_r - 2^{-\frac{2}{5}}[\hbar\frac{(1+2n)Coth(\frac{\hbar\omega}{k_B T})}{Cz_0^3\langle(\Delta i)^2\rangle L_l^{5/2}}]^{2/5}, \tag{14}$$

$$\mu = L_r - 2^{-\frac{2}{5}}[\hbar\frac{(1+2n)Coth(\frac{\hbar\omega}{k_B T})}{Cz_0^3\langle(\Delta i)^2\rangle C_l^{5/2}}]^{2/5} \tag{15}$$

where $n$ is the numbers of field photon corresponding to the number operator $\hat{n}$ in the Hilbert space.

## 3. Numerical simulations and discussions

Now we investigate the left-handedness of the mesoscopic CRLH-TL. The parameters used in our simulation are listed in Table 1. The order of magnitudes of these parameters are referenced from Ref.[32]. We consider one unit length of the mesoscopic CRLH-TL and set the quantum fluctuation of the current $\langle(\Delta i)^2\rangle = 25$, the field photon n=5 in Fig.2. As mentioned before, left-handedness means the simultaneously negative effective permittivity $\epsilon$ and permeability $\mu$ in the same frequency window. As shown in Fig.2, the effective permeability $\mu$ is negative in all frequency bands, as a result the mesoscopic CRLH-TL shows left-handedness in the frequency bands where the permittivity $\epsilon$ is negative, and shows right-handedness where the permittivity $\epsilon$ is positive.

Table 1: Parameters for the equivalent unit-cell circuit of the mesoscopic CRLH-TL.

|  | $C_l$ | $L_l$ | $C_r$ | $L_r$ | $\omega$ |
|---|---|---|---|---|---|
| Fig.2 | 148 $\mu F$ | 595 $\mu H$ | 35 $mF$ | 480 pH |  |
| Fig.3 | 250 $\mu F$ | 6.50 $mH$ | 9.0 $mF$ | 350 pH | 2.9GHz |
| Fig.4 | 550 $\mu F$ | 1.00 $mH$ | 45 $mF$ | 600 pH | 2.9GHz |

It's noted that the frequency ranges for negative permittivity $\epsilon$ increase when the temperatures enhance from 5K to 150K in Fig.2. And in the lower frequency bands, the negative permittivity $\epsilon$ are larger than those in the higher frequency band. What's more, the width of frequency bands for negative effective permittivity $\epsilon$ increases accompanying the temperature increase. We also notice that the left-handedness, i.e., the simultaneous values of negative effective permittivity $\epsilon$ and permeability $\mu$ can achieve surpassing the microwave frequency band, i.e. $\geq 3GHz$ when the temperature is higher than 85K. Fig.2 shows the left-handedness in the unit-cell circuit of the mesoscopic CRLH-TL succeeding in the lower frequency bands under a higher temperature.

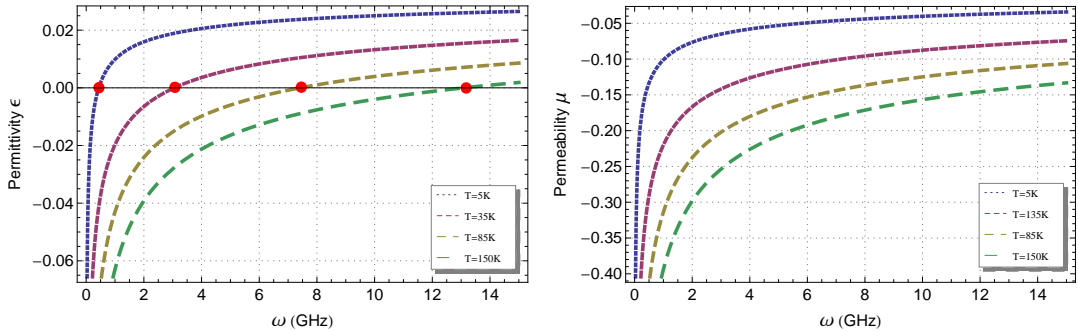

Figure 2: (Color online) The effective permittivity $\varepsilon$ and permeability $\mu$ of the mesoscopic CRLH-TL as a function of the frequencies via different temperatures: $T = 5K, T = 35K, T = 85K, T = 150K$.

The role of the intensity of current field in the mesoscopic CRLH-TL is one major factor for left-handedness. On account of the quantum Hamiltonian of the unit-cell circui is $\hat{H}=\hbar\omega(\hat{n}+\frac{1}{2})$, the intensity of the current field can be described by the numbers of the field photon. Profiting from Fig.2, the mesoscopic CRLH-TL operates in the microwave frequency band ($\omega = 2.9GHz$), and the quantum fluctuation of the current is set as $\langle(\Delta i)^2\rangle = 1$ in Fig.3. It is noted that the effective permittivity $\epsilon$ is negative when the temperature surpasses 170K with the weak current field (n=1). While the intensity of the current field increases gradually, the left-handedness can acquire at a lower temperature. The effective permittivity $\epsilon$ can be negative if the temperature surpasses 40K with a stronger current field (n=10). Fig.3 shows the fact that the mesoscopic CRLH-TL can be the left-handed material at a higher temperature when the weak current field travels through it, and at lower temperature with a more intense current field.

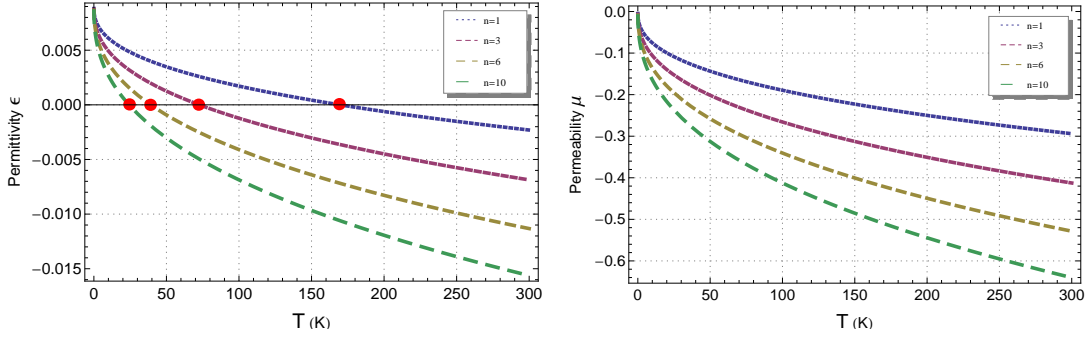

Figure 3: (Color online) The effective permittivity $\varepsilon$ and permeability $\mu$ of the mesoscopic CRLH-TL as a function of the temperatures under different numbers of the microwave field photon n: 1, 3, 6, 10.

The most distinctive quantum feature is the fluctuation of the current field in the mesoscopic CRLH-TL. The effect of the quantum fluctuation at different temperatures on the left-handedness is plotted in Fig.4. The effective permittivity $\varepsilon$ and permeability $\mu$ dependent the temperatures are shown by different the quantum fluctuations with the field photon n=1 in Fig.4. As shown in Fig.4, the effective permeability $\mu$ is negative in all temperature ranges, the left-handedness depends on the temperature ranges where the permittivity $\epsilon$ is negative. When the quantum fluctuation is weak $\langle(\Delta i)^2\rangle = 1.0$, a more wider temperature range for left-handedness is obtained. In other word, the left-handedness can arise from a low temperature to the room temperature when the fluctuation is weak in the mesoscopic CRLH-TL. However, the left-handedness can be achieved by the large number of the quantum fluctuation $\langle(\Delta i)^2\rangle = 4.5$ in a higher temperature environment.

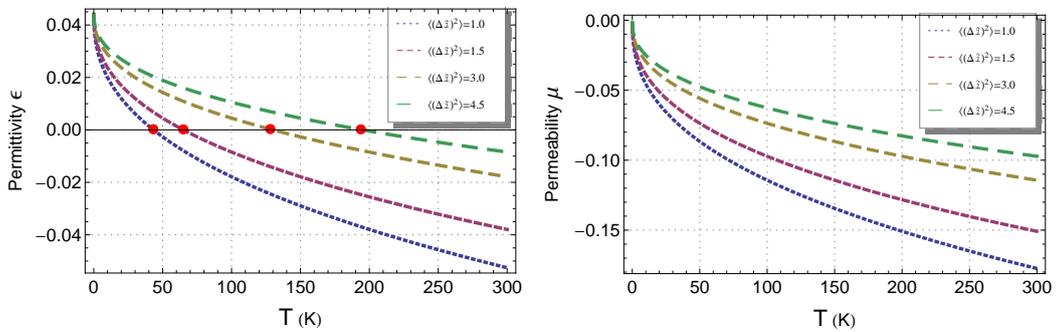

Figure 4: (Color online) The effective permittivity $\varepsilon$ and permeability $\mu$ of the mesoscopic CRLH-TL as a function of the temperatures under different the quantum fluctuation of the current $\langle(\Delta i)^2\rangle$: 1.0, 1.5, 3.0, 4.5.

Before concluding this paper we would like to point out that the frequency, intensity and fluctuations of the current field with different temperature in the mesoscopic CRLH-TL show the distinct characteristics in the left-handedness. We also remark that the discussion about the left-handedness doesn't relate to the quantum effect of the electronic components in the mesoscopic CRLH-TL, such as $C_l$ and $L_r$, $L_l$ and $C_r$. How to develop our present studies involving the quantum effect of the electronic components is to be considered in our forthcoming investigations.

## 4. Conclusion

In the present paper, we investigated the thermal effect on the left-handedness of the mesoscopic CRLH-TL, and the quantum thermal effect on the left-handedness can be summarized as: When the weak current field with a lower frequency and large quantum fluctuations travels through the mesoscopic CRLH-TL, the left-handedness can distinctly be achieved in the higher temperature environment. While the intense current field with lower frequency and weak fluctuations facilitates left-handedness at a lower temperature. For these reasons, we think the thermal effect on the left-handedness of the mesoscopic CRLH-TL deserves further experimental investigation in its miniaturization application.

**Acknowledgment**

This work is supported by the National Natural Science Foundation of China ( Grant Nos. 61205205 and 6156508508 ), the General Program of Yunnan Applied Basic Research Project, China ( Grant No. 2016FB009 ) and the Foundation for Personnel training projects of Yunnan Province, China ( Grant No. KKSY201207068 ).